# Vector solitons in $\mathcal{PT}$-symmetric lattices

Yaroslav V. Kartashov[*]

*Institute of Spectroscopy, Russian Academy of Sciences, Troitsk, Moscow Region, 142190, Russia*

I study vector solitons involving two incoherently-coupled field components in periodic $\mathcal{PT}$-symmetric optical lattices. The specific symmetry of the lattice imposes the restrictions on the symmetry of available vector soliton states. While all configurations with asymmetric intensity distributions are prohibited, such lattices support multi-hump solitons with equal number of "in-phase" or "out-of-phase" spots in two components, residing on neighboring lattice channels. In the focusing medium only the solitons containing out-of-phase spots in at least one component can be stable, while in the defocusing medium stability is achieved for structures consisting of in-phase spots. Mixed-gap vector solitons with components emerging from different gaps in the lattice spectrum also exist and can be stable in the $\mathcal{PT}$-symmetric lattice.

The concept of parity-time ($\mathcal{PT}$) symmetry has attracted considerable attention in different areas of science since the discovery of the fact that a specific class of complex potentials may have purely real spectrum in certain parameter range, provided that the shape $R(\eta)$ of this potential satisfies $\mathcal{PT}$-symmetry condition $R(\eta)=R^*(-\eta)$ [1,2]. Optical materials with inhomogeneous refractive index and gain/losses may allow simple physical realization of a $\mathcal{PT}$-symmetric system. The evolution of nonlinear optical modes in $\mathcal{PT}$-symmetric potentials was initially considered in [3,4], while various physical effects associated with non-orthogonality of Floquet-Bloch modes in $\mathcal{PT}$-symmetric lattices were discussed in [5]. It was realized that stable solitons in such systems exist for gain amplitudes below certain critical level at which the modes with complex eigenvalues appear and zero background becomes unstable. The experimental realization of the optical $\mathcal{PT}$-symmetric system was reported in [6].

Since then the properties of linear and nonlinear excitations were studied in various $\mathcal{PT}$-symmetric structures, including simple optical couplers [7-13], $\mathcal{PT}$-symmetric lattices with transverse refractive index gradients [14], binary and dimer structures [15-17], finite-dimensional structures [18], truncated [19] and two-dimensional [20] lattices, pseudo-potentials with $\mathcal{PT}$-symmetric nonlinear terms [21,22] and mixed linear-nonlinear lattices [23,24].

However, most of previous works address only scalar or single-field excitations in $\mathcal{PT}$-symmetric structures. The properties of vector states involving several coupled field components were not considered. At the same time, it is known that vector interactions considerably enrich the internal structure and stability properties of available soliton solutions in conservative settings [25-30].

Here I show that two incoherently coupled light fields propagating in the $\mathcal{PT}$-symmetric lattice may form stationary vector solitons composed of multiple light spots in both focusing and defocusing media. The stability of such states is determined by phase relations between neighboring spots. The increasing imaginary part of potential usually leads to destabilization of solitons in focusing medium, but stabilizes vector solitons in defocusing medium.

The propagation of two incoherently interacting light beams along the $\xi$-axis of the medium with transverse periodic modulation of the refractive index and gain/losses can be described by the system of coupled Schrödinger equations for the light field amplitude $q_{1,2}$ [30]:

$$i\frac{\partial q_{1,2}}{\partial \xi}=-\frac{1}{2}\frac{\partial^2 q_{1,2}}{\partial \eta^2}+\sigma q_{1,2}(|q_1|^2+|q_2|^2)-[p_r R_r(\eta)+ip_i R_i(\eta)]q_{1,2} \quad (1)$$

Here $\eta,\xi$ are the normalized transverse and longitudinal coordinates, respectively; the parameter $\sigma=+1(-1)$ corresponds to the defocusing (focusing) nonlinearity; $p_r$ is the normalized refractive index contrast, $p_i$ stands for the amplitude of gain/losses; the functions $R_r(\eta)$ and $R_i(\eta)$ describe the transverse profile of the refractive index and gain/losses, respectively. To satisfy the condition of $\mathcal{PT}$-symmetry we set $R_r(\eta)=\cos(\Omega\eta)$ and $R_i(\eta)=\sin(\Omega\eta)$, where the frequency $\Omega=4$ can be fixed by rescaling.

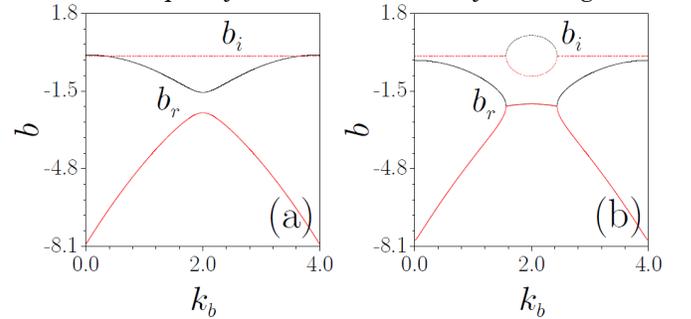

Fig. 1. Real part $b_r$ (solid curves) and imaginary part $b_i$ (dashed curves) of the propagation constant of Floquet-Bloch waves from the first and second allowed bands versus Bloch momentum $k_b$ at $p_i=0.5$ (a) and $p_i=2$ (b). In both cases $p_r=1$.

It is instructive to consider the properties of the spectrum of the complex potential $p_r R_r+ip_i R_i$ in linear medium ($\sigma=0$). The eigenmodes of such a potential are Bloch waves $w(\eta)\exp(ib\xi+ik_b\eta)$, which have the same periodicity as the underlying potential $w(\eta)=w(\eta+2\pi/\Omega)$. For a given value of Bloch momentum $k_b$ one obtains eigenvalues $b$ belonging to different allowed bands. The depend-ences $b(k_b)$ for the first two allowed bands are shown in Fig. 1. Despite the fact that the potential in Eq. (1) is complex, all eigenvalues $b$ remain real as long as the condition $p_i < p_r$ is satisfied [this implies that the corresponding Bloch waves propagate without catastrophic amplification or attenuation]. However, with the increase of $p_i$ the first finite gap shrinks and at $p_i > p_r$ first two allowed bands merge within certain interval of $k_b$ values [Fig. 1(b)]. Inside this region the eigenvalues have identical real parts $b_r$ and opposite imaginary parts $b_i$. Therefore, the corresponding Bloch waves will grow or decay upon propagation and zero background in Eq. (1) becomes unstable.

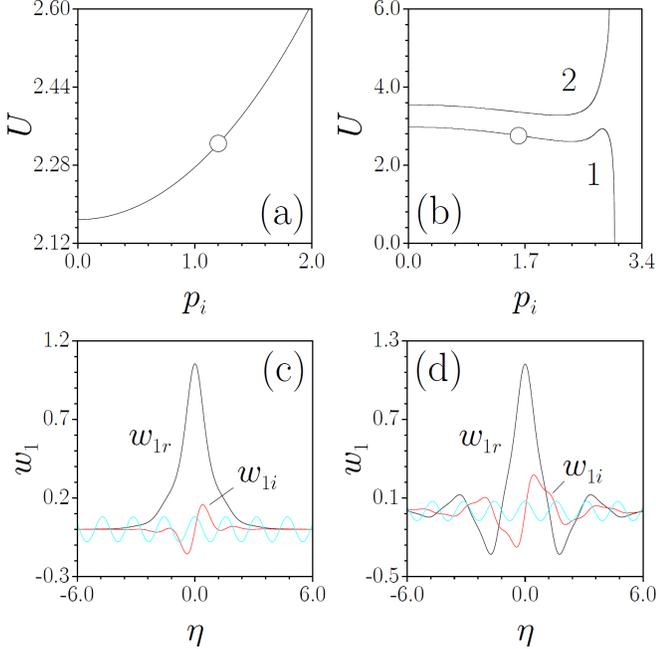

Fig. 2. (a) $U$ versus $p_i$ for odd-odd solitons in the focusing medium at $b_1 = b_2 = 1$, $p_r = 1$. (b) $U$ versus $p_i$ for odd-odd solitons in the defocusing medium at $b_1 = b_2 = -2$ (1) and $b_1 = b_2 = -2.3$ (2), for $p_r = 3$. Panels (c) and (d) show solitons corresponding to circles in (a) and (b), respectively. Cyan line shows the refractive index shape. Only the first component is shown.

Further I consider localized vector soliton solutions of Eq. (1) of the form $q_{1,2}(\eta,\xi) = w_{1,2}(\eta)\exp(ib_{1,2}\xi)$, where $w_{1,2}$ are the complex functions describing soliton shapes, and $b_{1,2}$ are the propagation constants belonging to different forbidden gaps in the linear spectrum of Fig. 1. I will fix the depth $p_r$ of the real part of potential and the propagation constant $b_2$ of one of the fields, and vary $b_1$ and $p_i$. It is well-known [26] that the case $b_1 = b_2$ is degenerated and corresponding solutions of Eq. (1) are of the form $w_1 = w\cos\phi$ and $w_2 = w\sin\phi$, where $\phi$ is the arbitrary projection angle, while $w$ is the solution of the scalar Schrödinger equation with the same potential. The properties of the simplest degenerated solutions with only one bright spot in each component (odd-odd solitons) are summarized in Fig. 2. In the focusing medium the total energy flow $U = U_1 + U_2 = \int_{-\infty}^{\infty}(|w_1|^2+|w_2|^2)d\eta$ of odd-odd soliton [Fig. 2(c)] from semi-infinite gap in the lattice spectrum monotonically grows with $p_i$ [Fig. 2(a)]. Such solitons are stable as long as zero background is stable at $p_i < p_r$. In the defocusing medium the simplest odd-odd solitons can be encountered in the first finite gap [Fig.

2(d)]. They possess oscillating tails typical for gap solitons. The dependence $U(p_i)$ for such solitons is non-monotonic [Fig. 2(b)]. For propagation constants close to the upper edge of the gap $U$ increases as $p_i \to p_r$ and at one point the tangential line to $U(p_i)$ dependence becomes vertical. In contrast, for propagation constants close to the lower gap edge the energy flow vanishes for sufficiently high $p_i$. Such solitons are stable close to the upper gap edge.

The most interesting situation is encountered in non-degenerated case, when $b_1 \neq b_2$. In this case soliton components should have different symmetries. Due to the fact that in $\mathcal{PT}$-symmetric landscapes the balance between gain and losses is very fragile, it can hardly be achieved for solitons with asymmetric intensity distributions. On this reason only the solutions featuring equal number of humps in each component were detected. A typical representative of this family is the twisted-even soliton, whose properties are described in Fig. 3 for the case of focusing medium. The components of such a soliton have propagation constants belonging to the semi-infinite gap in the lattice spectrum. While in the complex potential the solutions are chirped [Fig. 4(a)], at $p_i \to 0$ this family transforms into known family with out-of-phase spots in $w_1$ component and in-phase spots in $w_2$ component, residing on the neighboring refractive index maxima. Actually, by looking at the real part of the field of different components one can distinguish spots that can be conventionally termed "in-phase" and "out-of-phase" even at $p_i \neq 0$. Importantly, the increase of the imaginary part of potential strongly affects power sharing $S_{1,2} = U_{1,2}/U$ between components of vector soliton [Fig. 3(b)]. Upon increase of $p_i$ up to certain cutoff $p_i^{\mathrm{upp}}$ [indicated in Fig. 3(a)], the vector soliton transforms into scalar one, with all power concentrated either in $w_1$ [at large $b_1$ values as in Fig. 3(b)] or $w_2$ (at small $b_1$ values) component. The vector coupling between even $w_2$ component, that is unstable when propagating alone in the focusing medium, and stable twisted component $w_1$ results in stabilization of the entire vector complex. The latter is achieved for sufficiently large $b_1$ values close to the right edge of the existence domain on the plane $(b_1, p_i)$ [Fig. 3(d)]. Linear stability analysis shows that solitons are stable in the region below solid line marked with $p_i^{\mathrm{cr}}$ [this critical value is depicted in Fig. 3(a) and it was obtained from the dependence of the growth rate $\delta_r$ for weak perturbations on $p_i$ shown in Fig. 3(c)]. Notice that within a narrow range of propagation constants $p_i^{\mathrm{cr}}$ coincides with the threshold value $p_i = p_r$ [dashed line in Fig. 3(d)] at which the background is unstable. Further increase of $b_1$ leads to decrease of $p_i^{\mathrm{cr}}$ and shrinkage of the stability domain (as well as of the entire existence domain whose upper border is shown by the $p_i^{\mathrm{upp}}$ curve in the same figure). While stable twisted-even solitons retain their internal structure even for strong input perturbations, the unstable solitons from this family usually decay into odd-odd solitons if $p_i < p_r$.

Twisted-even solitons can be found not only in the focusing, but also in defocusing media. Their properties are described in Fig. 5, while representative profile is shown in Fig. 4(b). I consider vector solitons with both components emerging from the first finite gap, hence both $w_{1,2}$ functions feature pronounced oscillating tails due to Bragg reflection from periodic potential. As in the focusing medium the increase of the imaginary part of potential strongly affects the total power [Fig. 5(a)] and energy sharing [Fig. 5(b)] between soliton components. For small $b_1$ values close to the lower edge of the limited domain of soliton existence the vector complex transforms into scalar even soliton with all power concentrated in $w_2$ component. For large $b_1$ values close to the upper edge of the existence domain the dependences $S_{1,2}(p_i)$ become nonmonotonic as in the case shown in Fig. 5(b). Vector soliton complexes exist only below upper cutoff $p_i^{\mathrm{upp}}$ depicted with dashed line in Fig. 5(a). It is known [30,31] that in the defocusing medium only the solitons with in-phase spots in neighboring lattice periods can be stable. In vector case the stabilization of otherwise unstable twisted $w_1$ component can be achieved due to its coupling with stable even $w_2$ component. In complete contrast to the case of focusing nonlinearity, where growing gain-losses destabilize solitons, the increase of the depth of the imaginary part of potential in the defocusing medium results in stabilization of the composite vector state [Fig. 5(c) shows typical $\delta_r(p_i)$ dependence]. Stabilization occurs due to diminishing of the power fraction carried by twisted component with growth of $p_i$ [Fig. 5(b)]. Solitons become stable if $p_i$ exceeds certain critical value $p_i^{\mathrm{cr}}$. The domain of existence and stability of vector solitons in the defocusing medium is shown in Fig. 5(d). For selected parameters the upper edge of the existence domain $p_i = p_i^{\mathrm{upp}}$ is below the critical value $p_i = p_r$ depicted by dashed line. Solitons are stable in the domain $p_i^{\mathrm{cr}} < p_i < p_i^{\mathrm{upp}}$ that notably expands with increase of $b_1$. Examples of stable and unstable propagation of such modes are shown in Fig. 6.

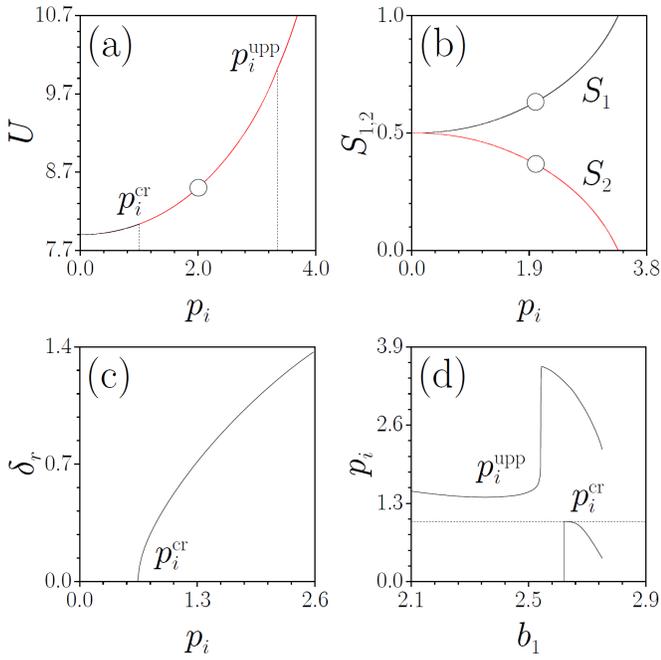

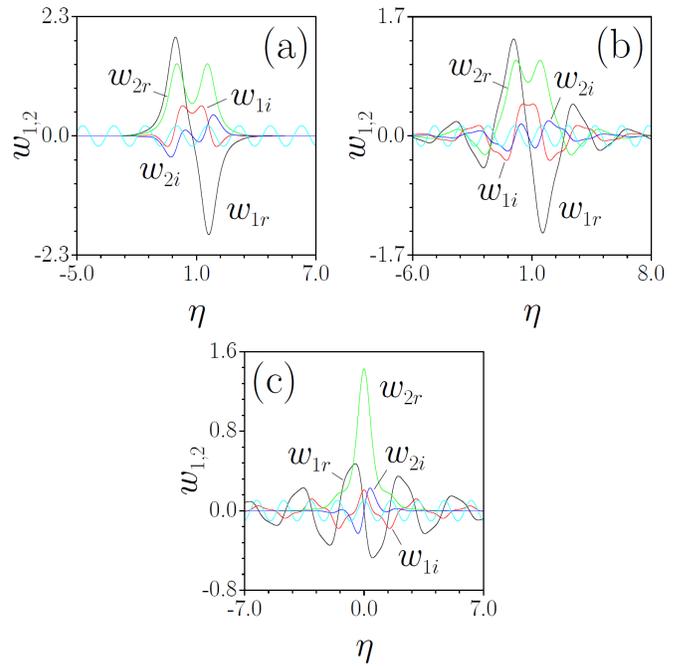

Fig. 3. Energy flow (a) and energy sharing between components of the twisted-even vector soliton (b) in the focusing medium versus $p_i$ at $b_1 = 2.62$, $b_2 = 3$. Circles correspond to soliton shown in Fig. 4(a). (c) $\delta_r$ versus $p_i$ at $b_1 = 2.72$, $b_2 = 3$. (d) The domain of existence and stability on the plane $(b_1, p_i)$ at $b_2 = 3$. Dashed line shows the critical value $p_i \equiv p_r$. In all cases $p_r = 1$.

Fig. 4. Profiles of (a) twisted-even soliton in the focusing medium at $p_i = 2$, $p_r = 1$, $b_1 = 2.62$, $b_2 = 3$, (b) twisted-even soliton in the defocusing medium at $p_i = 1.5$, $p_r = 3$, $b_1 = -2.66$, $b_2 = -2$, (c) and odd-odd soliton in the focusing medium at $p_i = 1.8$, $p_r = 3$, $b_1 = -2.9$, $b_2 = 1.5$.

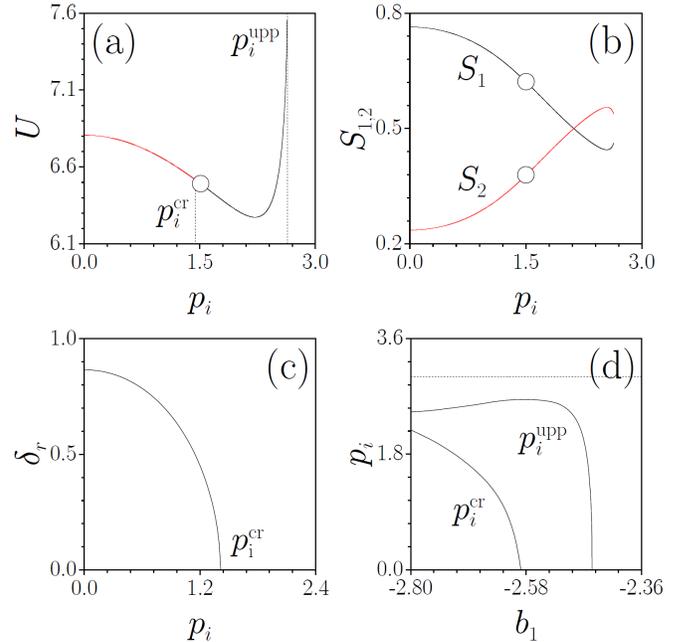

Fig. 5. Energy flow (a), energy sharing (b), and perturbation growth rate (c) versus $p_i$ for the twisted-even soliton in the defocusing medium at $b_1 = -2.66$, $b_2 = -2$. Circles correspond to soliton from Fig. 4(b). (d) The domain of existence and stability on the plane $(b_1, p_i)$ at $b_2 = -2$. Dashed line indicates the critical value $p_i \equiv p_r$. In all cases $p_r = 3$.

Besides simplest twisted-even solitons described above one may obtain more complex soliton families containing more than two bright spots in each field component in both focusing and defocusing media. Such solitons also can be stable for properly selected phase distributions.

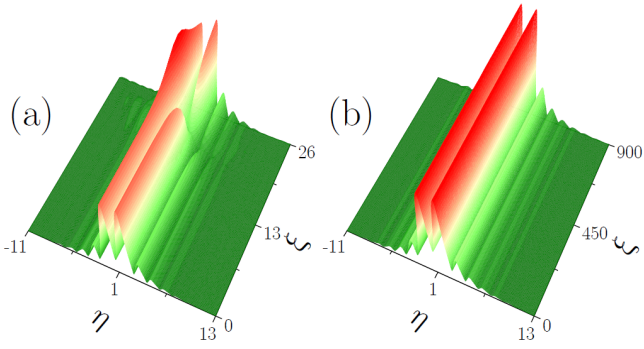

Fig. 6. Decay of unstable twisted-even soliton at $p_i=1.1$ (a) and its stable propagation at $p_i=1.5$ (b) in defocusing medium. In both cases $p_r=3$, $b_1=-2.66$, $b_2=-2$.

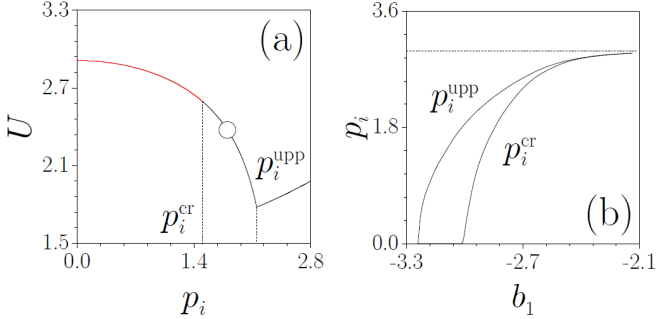

Fig. 7. (a) $U(p_i)$ curve for mixed-gap soliton in the focusing medium at $b_1=-2.9$, $b_2=1.5$, $p_r=3$. Circle corresponds to soliton in Fig. 4(c). (b) The domain of existence and stability on the plane $(b_1,p_i)$ at $b_2=1.5$, $p_r=3$. Dashed line corresponds to $p_i\equiv p_r$.

Finally, vector solitons may appear due to coupling of components whose propagation constants belong to different gaps in the lattice spectrum [27,28]. Here I consider such states in $\mathcal{PT}$-symmetric lattice in the case of focusing nonlinearity, when propagation constants of soliton constituents belong to the semi-infinite ($b_2$) and first finite ($b_1$) gaps. A representative profile of mixed-gap vector soliton is shown in Fig. 4(c). While its second component from the semi-infinite gap is bell-shaped, the first component possesses multiple field oscillations. Different symmetries of the components reflect the fact that they belong to different gaps. For sufficiently large $b_1$ values, close to the right edge of the existence domain in Fig. 7(b), increasing $p_i$ leads to the transformation of vector soliton into the scalar one, with all power concentrated in the component from the first finite gap. The upper edge of the existence domain of vector states at $p_i=p_i^{\mathrm{upp}}$ is indicated by dashed line in Fig. 7(a) (the scalar family remains beyond this point). The first component that is unstable when it propagates alone can be stabilized due to coupling with stable second component from the semi-infinite gap. Stable solitons exist for $p_i^{\mathrm{cr}}<p_i<p_i^{\mathrm{upp}}$ and domain of stability broadens with decrease of propagation constant $b_1$ [Fig. 7(b)]. As in the case of defocusing nonlinearity the existence domain of mixed-gap solitons is located below $p_i=p_r$ line, corresponding to the appearance of complex eigenvalues in the spectrum of $\mathcal{PT}$-symmetric potential.

Summarizing, $\mathcal{PT}$-symmetric lattices support a variety of stable vector solitons consisting of multiple bright spots in both focusing and defocusing media. The imaginary part of $\mathcal{PT}$-symmetric drastically affects the internal structure and stability of such states.